\documentclass[runningheads]{llncs}
\usepackage{graphicx}
\usepackage{graphicx}
\usepackage[square,numbers]{natbib}

\begin{document}

\title{Empirical Study of Quality Image Assessment for Synthesis of Fetal Head Ultrasound Imaging with DCGANs}

\author{
T. Bautista \and 
J. Matthew \and 
H. Kerdegari \and 
L. Peralta \and 
M. Xochicale 
}
\authorrunning{Bautista et al.}

\institute{
King's College London, \\ 
School of Biomedical Engineering \& Imaging Sciences, \\ 
London SE1 7EU \\
\email{\{thea.bautista, miguel.xochicale\}@kcl.ac.uk} 
}
\maketitle              

\begin{abstract}
In this work, we present an empirical study of DCGANs, including hyperparameter heuristics and image quality assessment, as a way to address the scarcity of datasets to investigate fetal head ultrasound.
We present experiments to show the impact of different image resolutions, epochs, dataset size input, and learning rates for quality image assessment on four metrics: mutual information (MI), Fr\'echet inception distance (FID), peak-signal-to-noise ratio (PSNR), and local binary pattern vector (LBPv). 
The results show that FID and LBPv have stronger relationship with clinical image quality scores.
The resources to reproduce this work are available at 
\url{https://github.com/budai4medtech/miua2022}.
\end{abstract}

\section{Introduction}
\begin{figure}
\includegraphics[width=\textwidth]{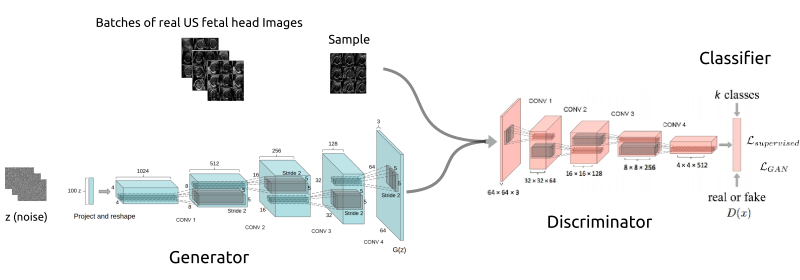} 
\caption{DCGAN architecture for US imaging. Figure is adapted from the work of Radford et al. 2016 \cite{radford2016unsupervised}.} \label{fig1}
\end{figure}

Synthesis of Ultrasound (US) imaging is growing in medical communities as a way to mitigate the scarcity of datasets due to the cost of data collection, annotation and its ethical policies with a generalisable and reproducible deep learning pipeline \cite{fiorentino2022-DLUSFetalImaging}.
Generative Adversarial Networks (GANs) have the ability to mimic data distributions and synthesise medical images \cite{KAZEMINIA2020gans, skandarani2021gans}.
In the case of US imaging, GANs are able to synthesise 2D images with fetal phantoms \cite{hu2017-gans-us}, intravascular US \cite{tom-gans-us}, breast US \cite{fujioka2019-gans-us}, and fetal head US \cite{lee2022-gans-us}.
Recently, Skandarani \textit{et al.} in 2021 presented an empirical study for GANs in medical imaging on the impact of sensitivity of hyperparameters, dataset and computer scale, FID and image quality, and its clinical usability \cite{skandarani2021gans}.
However, there are little to no empirical studies on the use of GANs with US fetal imaging but nerve US \cite{kumar2021empirical}.
Considering that Deep Convolutional GANs (DCGANs) improve image quality generation and training stability of the networks (stride CNNs, extensive use of batchNorm, etc) \cite{radford2016unsupervised} and its wide application in different medical image modalities \cite{radford2016unsupervised,KAZEMINIA2020gans, skandarani2021gans, fujioka2019-gans-us}, we apply DCGAN architecture \cite{radford2016unsupervised} with different input pixel image size (e.g. DCGAN64 and DCGAN128) (Fig.~\ref{fig1}) using an open dataset of 999 real fetal head US images \cite{vandenHeuvel2018}.

\section{Quality assessment of synthetic fetal head US imaging}
For quality assessment of synthetic fetal head US imaging, we consider four metrics for image quality assessment: The Fr\'echet Inception Distance (FID), peak-signal-to-noise ratio (PSNR), mutual information (MI), and local binary pattern vector (LBPv). 
FID is one of the most widely used metrics for GANs, where a lower FID indicates better image quality and increased diversity \cite{heusel2017}.
PSNR is a measurement of the ratio between the maximum power of a signal and the difference between the original and synthetic image, where a higher PSNR indicates closer intensity similarity between the synthetic and original US images. 
MI is a measure of image similarity and indicates the mutual dependence between two images, where higher MI denotes greater similarity between the images \cite{wang2019}.
LBPv are texture descriptors, where the closer these values are from a synthetic image to a reference LBPv calculated from real US images, the higher the image quality is \cite{singh2018}. 

\section{Experiments and results}
\begin{figure}
\includegraphics[width=\textwidth]{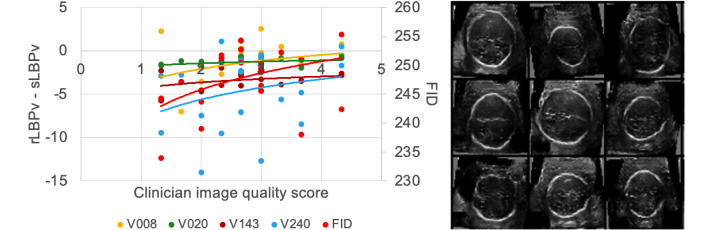} 
\caption{Regression plots for FID, V8, V20, V143 and V240 with clinician image quality scores (left) and nine synthetic images of 64x64 pixel size from DCGAN64 trained on 300 images after 300 epochs (right).}\label{fig2}
\end{figure}

Fetal head US images were resized to 64x64 (due to training instability at higher resolutions) and augmentation was applied in the form of random horizontal flipping and rotation to mimic variation in US probe location.
Then DCGAN model was trained with the original hyperparameters, (learning rate=0.0002; optimiser=Adam optimiser; $\beta1=0.5$; $\beta2= 0.999$ and loss function=binary cross entropy loss  \cite{radford2016unsupervised}), demonstrating the most stable training and resulting in the most realistic synthetic images when training the model on augmented data.
Hence, the model was trained on 100, 300, 500, 800 and 999 US images, creating 20 sets of 800 synthetic images with 300, 500, 800 and 1000 epochs.
FID, MI, PSNR and the LBPv values (sLBPv) were calculated using these sets of synthetic images.
A reference LBPv (rLBPv) was calculated from 800 non-augmented original images from which the sLBPv was compared to.
To assess the usefulness of FID, PSNR, MI and LBPv, a clinician with 10 years' experience with fetal US and two biomedical engineers were asked to differentiate between 68 real and synthetic images and to rate image quality on a scale of 1-5 (1 being poor quality and 5 being high quality).
Image quality in the generated images did not vary significantly with increasing training size (MI, PSNR, FID and LBPv did not differ significantly with $p>0.05$) which could indicate an upper limit in the capability of DCGAN in generating synthetic fetal US images.
The biomedical engineers indicated 56.14\% of the synthetic images as 'real' which shows that the DCGAN was able to create images that were indifferentiable from the original ones.
However, the clinician had marked all images in the survey as 'fake' due to the small resolution and some instances of mode collapse reduced variation in synthetic images; both of which mitigate clinical application.
FID and the V8, V20, V143 and V240 elements of the LBPv showed the strongest relationship to clinician image quality scores in regression analysis and highest correlation scores ($r>0.2$) in comparison to the other metrics (Fig.~\ref{fig2}, left); the small correlation was possibly attributed to the low number of data points used for analysis. These LBPv elements also had the smallest difference to the rLBPv thus show promise as indicators for textural similarity.

\section{Conclusions and future work}
To conclude, DCGANs are capable of satisfactorily learning the data distribution of the training images so as to create synthetic fetal US images indifferentiable from the originals, where FID and LBPv metrics show potential for image quality assessment of synthetic fetal US in comparison to MI and PSNR.
However, the synthesised images are limited in clinical application due to their resolution of 64x64 (less than 500x500 pixels) and artefacts such as mode collapse.
Hence, to create higher resolution US images for clinical applications, future work might lead to testing other GAN models, architectural alterations, hyperparameter optimization, addresing mode collapse artefacts and using other forms of data augmentation with other fetal US datasets.



\bibliographystyle{apalike}


\end{document}